 \documentclass[12pt,preprint]{emulateapj}
 \usepackage{graphicx}

\newcommand{\ang}{{\rm \AA}}

\newcommand{\etal}{{\it et~al.\/}}

\newcommand{\Hline}[1]{\mbox{H{\footnotesize {#1}}}}
\newcommand{\Halpha}{\Hline{\mbox{$\alpha$}}}

\newcommand{\HI}{{\sc Hi}}
\newcommand{\HII}{{\sc Hii}}
\newcommand{\HIMF}{{\sc HiMF}}
\newcommand{\HICAT}{{\sc HiCAT}}
\newcommand{\HIPASS}{{\sc HiPASS}}

\newcommand{\lha}{\mbox{$l_{\rm H\alpha}$}}
\newcommand{\lhap}{\mbox{$l'_{\rm H\alpha}$}}
\newcommand{\Lsun}{\mbox{$L_\odot$}}
\newcommand{\MHI}{\mbox{${\cal M}_{\rm HI}$}}
\newcommand{\MHtwo}{\mbox{${\cal M}_{\rm H2}$}}

\newcommand{\Mdyn}{\mbox{${\cal M}_{\rm dyn}$}}
\newcommand{\Msun}{\mbox{${\cal M}_\odot$}}
\newcommand{\Mstar}{\mbox{${\cal M}_\star$}}
\newcommand{\NII}{{\sc Nii}}
\newcommand{\rHII}{\mbox{$r_{\rm HII}$}}
\newcommand{\rmax}{\mbox{$r_{\rm max}$}}
\newcommand{\rhoHI}{\mbox{$\rho_{\rm HI}$}} 
\newcommand{\SFRD}{\mbox{$\dot{\rho}_{\rm SFR}$}} 
\newcommand{\SFRDo}{\mbox{$\dot{\rho}_{\rm SFR}(0)$}} 
\newcommand{\SFRDunits}{\([\Msun\ \mbox{\rm yr}^{-1}\ \mbox{\rm Mpc}^{-3}]\)}
\newcommand{\SFRDp}{\mbox{$\dot{\rho}'_{\rm SFR}$}} 
\newcommand{\SFRDpo}{\mbox{$\dot{\rho}'_{\rm SFR}(0)$}} 
\newcommand{\SFRDz}{\mbox{$\dot{\rho}_{\rm SFR}(z)$}} 
\newcommand{\SFRDpz}{\mbox{$\dot{\rho}'_{\rm SFR}(z)$}} 

\newcommand{\tgas}{\mbox{$t_{\rm gas}$}}
\newcommand{\torb}{\mbox{$t_{\rm orb}$}}

\slugcomment{Submitted to ApJ 2006 January 3, Accepted 2006 March 20}

\shorttitle{SINGG. II. Local Star Formation Rate Density}
\shortauthors{Hanish \etal }

\begin{document}

\title{The Survey for Ionization in Neutral Gas Galaxies- II. The Star Formation Rate Density of the Local Universe}

\author{D.J.\ Hanish\altaffilmark{1},
G.R.\ Meurer\altaffilmark{1},
H.C.\ Ferguson\altaffilmark{2},
M.A.\ Zwaan\altaffilmark{3},
T.M.\ Heckman\altaffilmark{1},
L.\ Staveley-Smith\altaffilmark{4},
J.\ Bland-Hawthorn\altaffilmark{5}, 
V.A.\ Kilborn\altaffilmark{6,4},
B.S.\ Koribalski\altaffilmark{4},
M.E.\ Putman\altaffilmark{7},
E.V.\ Ryan-Weber\altaffilmark{8},
M.S.\ Oey\altaffilmark{7},
R.C.\ Kennicutt, Jr.\altaffilmark{9},
P.M.\ Knezek\altaffilmark{10},
M.J.\ Meyer\altaffilmark{3},
R.C.\ Smith\altaffilmark{11},
R.L.\ Webster\altaffilmark{12},
M.A.\ Dopita\altaffilmark{13},
M.T.\ Doyle\altaffilmark{14},
M.J.\ Drinkwater\altaffilmark{14},
K.C.\ Freeman\altaffilmark{13},
J.K.\ Werk\altaffilmark{7}}

\altaffiltext{1}{Department of Physics and Astronomy, The Johns Hopkins University, 3400 North Charles St., Baltimore, MD 21218-2686, $hanish@pha.jhu.edu$}
\altaffiltext{2}{Space Telescope Science Institute, 3700 San Martin Drive, Baltimore, MD 21218}
\altaffiltext{3}{European Southern Observatory, Karl-Schwarzschild-Str. 2, D-85748 Garching b. M\"unchen, Germany}
\altaffiltext{4}{Australia Telescope National Facility, CSIRO, P.O. Box 76, Epping, NSW 1710, Australia.}
\altaffiltext{5}{Anglo-Australian Observatory, P.O. Box 296, Epping, NSW 2121, Australia}
\altaffiltext{6}{Centre for Astrophysics and Supercomputing, Swinburne University of Technology, Mail 39, P.O. Box 218, Hawthorn, VIC 3122, Australia}
\altaffiltext{7}{Department of Astronomy, University of Michigan, Ann Arbor, MI 48109}
\altaffiltext{8}{Institute of Astronomy, University of Cambridge, Madingley Road, Cambridge, CB3 0HA, UK}
\altaffiltext{9}{Steward Observatory, University of Arizona, Tucson, AZ 85721}
\altaffiltext{10}{WIYN, Inc., 950 North Cherry Ave., Tucson, AZ 85726}
\altaffiltext{11}{Cerro Tololo Inter-American Observatory (CTIO), Casilla 603, La Serena, Chile}
\altaffiltext{12}{School of Physics, University of Melbourne, Parkville, VIC 3010, Australia}
\altaffiltext{13}{Research School of Astronomy and Astrophysics (RSAA), Australian National University, Cotter Road, Weston Creek, ACT 2611, Australia}
\altaffiltext{14}{Department of Physics, University of Queensland, Brisbane, QLD 4072, Australia}

\begin{abstract}

We derive observed \Halpha\ and $R$ band luminosity densities of an
\HI -selected sample of nearby galaxies using the SINGG sample to be
$\lhap = (9.4 \pm 1.8) \times 10^{38}\ h_{70}$ erg s$^{-1}$ Mpc$^{-3}$
for \Halpha\ and $l_R' = (4.4 \pm 0.7) \times 10^{37}\ h_{70}$ erg
s$^{-1}$ \ang$^{-1}$ Mpc$^{-3}$ in the $R$ band.  This $R$ band
luminosity density is approximately 70\% of that found by the Sloan
Digital Sky Survey.  This leads to a local star formation rate density
of $\log$(\SFRD\ \SFRDunits) = $-1.80\ ^{+0.13}_{-0.07}(random)\ \pm
0.03(systematic) + \log(h_{70})$ after applying a mean internal
extinction correction of 0.82 magnitudes.  The gas cycling time of
this sample is found to be \tgas\ = $7.5\ ^{+1.3}_{-2.1}$ Gyr, and the
volume-averaged equivalent width of the SINGG galaxies is $EW(\Halpha
) = 28.8\ ^{+7.2}_{-4.7}\ $\ang$\ (21.2\ ^{+4.2}_{-3.5}\ $\ang\ without
internal dust correction).  As with similar surveys, these results
imply that \SFRDz\ decreases drastically from $z \sim 1.5$ to the
present.  A comparison of the dynamical masses of the SINGG galaxies
evaluated at their optical limits with their stellar and \HI\ masses
shows significant evidence of downsizing: the most massive galaxies
have a larger fraction of their mass locked up in stars compared with
\HI , while the opposite is true for less massive galaxies.  We show
that the application of the Kennicutt star formation law to a galaxy
having the median orbital time at the optical limit of this sample
results in a star formation rate decay with cosmic time similar to
that given by the \SFRDz\ evolution.  This implies that the \SFRDz\
evolution is primarily due to the secular evolution of galaxies,
rather than interactions or mergers.  This is consistent with the
morphologies predominantly seen in the SINGG sample.

\end{abstract}


\section{Introduction} \label{s:intro}

The star formation rate density of the universe has changed
considerably since $z \sim 2$, decreasing by approximately an order of
magnitude.  This decrease has been widely discussed
\citep[e.g.][]{b:madau96, b:pei99, b:somerville01, b:hopkins04}
because the evolution of the star formation rate density acts to
constrain all models of galaxy formation and evolution.
Redshift-dependent luminosity densities (such as the $R$ band
luminosity density $l_R(z)$) and star formation rate densities (\SFRDz
) remain some of the best constraints on these models.  The value of
$l_R(z)$ at $z \approx 0$ constrains the evolution of stellar mass
\citep{b:madau98}.  Likewise, the value of \SFRDz\ at $z \approx 0$
helps determine the relative contributions of burst and quiescent star
formation \citep{b:somerville01} and the chemical evolution of the
universe \citep{b:pei95}.  Estimates of \SFRD\ from previous studies
span a factor of two or more, and do not always agree within their
stated uncertainties, as shown in Table \ref{t:sfrd}.  All surveys
suffer biases, however, and these may explain the large discrepancies
between the densities derived from each.

The process of star formation leaves measurable signatures across the
electromagnetic spectrum, allowing numerous methods for selecting
star-forming galaxies, each with its own set of biases.  For example,
objective prism surveys for emission line galaxies
\citep[e.g.][]{b:gallego95, b:gronwall97} result in a large,
consistent bias towards galaxies with high surface brightness, high
equivalent width emission lines \citep[e.g.][]{b:salzer89}.
Ultraviolet (UV) surveys \citep[e.g.][]{b:treyer98} are biased against
galaxies highly attenuated by dust.  Conversely, far-infrared (FIR)
\citep[e.g.][]{b:flores99, b:perez05} or sub-millimeter surveys select
galaxies by the presence of dust-reprocessed optical and UV light,
resulting in a bias against galaxies possessing little dust.
Similarly, 1.4 GHz radio surveys \citep[e.g.][]{b:serjeant02} bias
against systems with low dust content \citep{b:bell03}; in addition,
the correlation between radio flux and FIR flux (which is then used to
calculate star formation rate) is not linear, resulting in a
pronounced bias against low luminosity galaxies
\citep{b:devereux89,b:yun01}.

The 21-cm spectral line of neutral hydrogen (\HI ) presents a useful
starting point for star formation surveys.  New stars form out of the
interstellar medium, of which \HI\ is a key component.  While it is
the molecular component of the ISM from which new stars form, a strong
correlation has been found between \HI\ surface density and star
formation intensity \citep{b:kennicutt98}.  \HI\ flux is unaffected by
dust extinction, and has been measured in every type of star-forming
galaxy.  As a result, selecting targets by \HI\ mass creates a sample
set free of the common optical biases.  For this, we use the \HI\
Parkes All Sky Survey (\HIPASS ), an \HI\ survey of the southern sky
($\delta \le +2^{\circ}$) over a velocity range from $-1280$ to +12700
km s$^{-1}$ obtained with the 21-cm multibeam receiver
\citep{b:staveley96} at the 64m Parkes radio telescope
\citep{b:barnes01}.


We have commenced the Survey of Ionization in Neutral Gas Galaxies
(SINGG) to survey star formation in the \HIPASS\ galaxies.
\citet[][hereafter Paper I]{b:meurer06} introduce the SINGG survey,
including sample selection, methods used, and basic measurements.  In
total the SINGG sample set includes 468 \HI -selected targets covering
the mass range $7.0 \le \log(\MHI/\Msun ) \le 11.0$ in approximately
equal numbers per decade of mass.  Distance was also used in the
selection process; the nearest sources were preferentially selected at
each mass to improve physical resolution and avoid confusion.

To measure star formation, SINGG uses \Halpha, the most readily
accessible optical tracer of star formation at low redshifts.
\Halpha\ is a recombination line primarily resulting from hydrogen
photoionization.  Because this requires ionizing UV photons, the
majority of \Halpha\ flux will be produced near the most massive O
stars, whose extremely short lifespans ($\lesssim 10$ Myr) make them
good indicators of the current star formation rate.  SINGG uses $R$
band measurements for continuum subtraction, which also provide useful
measurements of the existing stellar populations.  Observations
presented in Paper I consist of the 93 \HIPASS\ extragalactic \HI\
targets fully processed to date.  Due to the large beam size of
\HIPASS, 13 of these \HI\ targets contain between two and four
distinct \Halpha\ objects.  As a result, the SINGG sample includes a
total of 111 individual \Halpha -emitting galaxies, which we refer to
as SINGG Release 1, or SR1.  We however exclude one \HI\ target,
J0403-01, and its single \Halpha -emitting galaxy due to excessive sky
uncertainties and foreground field contamination.  The results
included in this paper are entirely generated from the data presented
in Paper I.

This paper uses the SINGG observations to derive the \Halpha\ and $R$
band luminosity densities of the local universe, designated $\lha(z)$
and $l_R(z)$ respectively, where $z$ indicates the mean redshift of
the survey.  For this and most of the other local ($z \approx 0$)
studies, we will omit the $(z)$ notation except where required.
Combined with the \HI\ data from \HIPASS , these yield the star
formation rate density of the local universe, the stellar luminosity
density, the density of the neutral ISM, and the cosmic gas cycling
time.  Section~\ref{s:method} explains the methodology used to
determine the various volume densities and their uncertainties.
Section~\ref{s:results} gives the results of our calculations.
Section~\ref{s:disc} gives the results and compares to those of other
surveys, while Section~\ref{s:concl} discusses some of the
implications of our results.  We use a $\Lambda$CDM cosmology
($\Omega_{0}$ = 0.3, $\Omega_{\Lambda}$= 0.7) with a Hubble constant
of $H_{0}$ = 70\ km s$^{-1}$ Mpc$^{-1}$, and a \citet{b:salpeter55}
IMF between 0.1 and 100 \Msun .

\section{Methodology} \label{s:method}

\subsection{\HI\ mass}

Since the SINGG sample is not volume complete, we tie our results to
an \HI\ Mass Function (\HIMF).  \HI\ parameters are primarily derived
from the data given in our two main source catalogs: \HICAT , the
final \HIPASS\ catalog \citep{b:meyer04}, and BGC, the \HIPASS\ Bright
Galaxy Catalog \citep{b:koribalski04}.  These catalogs have 95\%\
completeness limits in velocity-integrated 21 cm flux densities of
$\int S_{\nu}\ d\rm v \sim $ 5 and 25 Jy km s$^{-1}$, respectively.
The \HI\ mass of a galaxy at a distance of $D$ Mpc is derived using
the standard relation $\MHI\ [\Msun ]= 2.36 \times 10^{5} D^{2} \int
S_{\nu}\ d$v [Jy km s$^{-1}$] \citep{b:roberts62}.  As a result, the
flux density limits for these two samples correspond to \HI\ mass
limits of 1.2 and 5.9 $\times\ 10^{6}\ D^{2}\ \Msun $, respectively.
The SR1 sources have distances of 4 -- 73 Mpc (with the majority
falling within the 10 -- 20 Mpc range) and \HI\ masses of $10^{7.5}$
-- $10^{10.6}$ \Msun .  Detailed information about the \HI\ masses of
the SINGG galaxies can be found in the original catalog publications
and Paper I.

Distances for most sources are derived from radial velocities using
the model of \citet{b:mould00}, which corrects for infalls towards the
Virgo cluster, Great Attractor, and Shapley supercluster.  Distances
to the nearest galaxies are taken from \citet{b:karachentsev04}.  This
alters the distances and \HI\ masses for many SINGG galaxies when
compared to other \HIPASS\ papers \citep{b:zwaan05}, and marginally
alters the \HI\ Mass Function used to correct for incompleteness.  The
values given throughout this paper are those derived using the Mould
model for the full \HIPASS\ sample.

We use the standard Schechter function to parameterize the \HIMF:
\begin{equation}
\theta(\MHI )\ d\MHI\ = \theta_{\ast} \left( \frac{\MHI }{\Mstar
} \right)^{\alpha} e^{- \left( \frac{\MHI }{\Mstar } \right) }\
d\left( \frac{\MHI }{\Mstar } \right)
\end{equation}
where $\theta(\MHI )$ represents the number density of galaxies as a
function of \HI\ mass (in Mpc$^{-3}$), \Mstar\ is the characteristic
mass, $\alpha$ is the ``faint'' end slope and $\theta_\ast$ is the
normalization factor.  We calculate Schechter fits to raw, binned
$\theta(\MHI )$ data from \citet{b:zwaan05}, and adjusted to the
distance of \citet{b:mould00} for $H_0 = 70\, {\rm km\ s^{-1}\
Mpc^{-1}}$.  This yields $\Mstar = 10^{9.92 \pm 0.04}\ h_{70}^{-2}$
\Msun , $\alpha = -1.41 \pm 0.05$ and $\theta_{\ast} = (3.86 \pm 0.7)
\times 10^{-3}\ h_{70}^{3}$ Mpc$^{-3}$ dex$^{-1}$; these values are
compared to other \HI\ Mass Functions in Table~\ref{t:himf}.  As is
usually the case, the errors on \Mstar\ and $\theta_{\ast}$ are highly
correlated.

For the 13 SR1 \HI\ targets containing multiple \Halpha\ sources,
\HIPASS\ can only provide the total \HI\ mass for each target, with no
ability to distinguish the contributions of each individual galaxy.
When calculating our luminosity densities, the luminosities of the
individual galaxies within each \HI\ target are combined and the total
is treated as a single aggregate object.  As the mass function given
above was generated using similarly combined \HI\ masses, this
approach should not substantially bias our results.

\subsection{Volume densities} \label{s:calc}

Using the \HIMF , we derive the $R$ continuum luminosity density
$l_R$, \Halpha\ luminosity density \lha , and \HI\ mass density
\rhoHI\ for the local universe.  We denote values uncorrected for
internal extinction with a prime ($'$) symbol.

The volume density of a quantity $x$ is found using:
\begin{equation}
n_{x} = \int\ \theta (\MHI )\ x(\MHI ) \
d\left(\frac{\MHI}{\Mstar}\right)
\label{e:rhox}
\end{equation}
where $x = L_{\rm H\alpha}$ when calculating \lha, $x = L_R$ when
calculating $l_R$, $x = 1.0$ when calculating the number density $n$,
or $x = \MHI$ when calculating \rhoHI .

By separating our data into \HI\ mass bins and combining the results
within each bin, Eq.~\ref{e:rhox} is replaced by:
\begin{equation}
n_{x} = \ln (10) \sum_{i=1}^{N_{bins}} \frac{\Delta
\log(\MHI)_{i}}{N_{i}} \sum_{j=1}^{N_{i}} \theta (\MHI_{j})\
x_{j} \left(\frac{\MHI_{j}}{\Mstar}\right)
\label{e:rhox2}
\end{equation}
Here, $i$ represents the array of mass bins, while $j$ represents the
individual \HIPASS\ targets within each bin.  $\Delta \log(\MHI)_{i}$
is the logarithmic width of each mass bin; our bins are 0.5 decades
wide for all bins except the lowest ($7.0 \le \log(\MHI/\Msun ) \le
8.0$), as shown in Table~\ref{t:bindata}.

The SR1 \Halpha\ and $R$ luminosities are given in Paper I, along with
lists of targets containing multiple emission line galaxies.  These
luminosities are derived from fluxes extracted using elliptical
apertures, supplemented with the flux from outer disk \HII\ regions.
Full discussions of our $R$ and \Halpha\ flux extraction procedures
are given in Paper I.

To calculate the Star Formation Rate (SFR) for each galaxy, we adopt
the conversion from \Halpha\ luminosity to SFR given by
\citet{b:kennicutt94}.  This relationship is derived using a single
power law Initial Mass Function (IMF) having a \citet{b:salpeter55}
slope and spanning the mass range of 0.1 to 100 \Msun .  The resulting
conversion is
\begin{equation}
SFR\ [\Msun\ \mbox{\rm yr}^{-1}] = \left(\frac{L_{\rm H\alpha}\ [\rm
erg\ s^{-1}]}{1.26 \times 10^{41}}\right)
\label{e:sfr}
\end{equation}
We adopt this conversion to maintain consistency with numerous other
studies.  The choice in IMF is a major source of systematic
uncertainty, since the conversion of $L_{\rm H\alpha}$ to SFR depends
entirely on the fraction of stars falling within the mass range
responsible for ionizing hydrogen.  For example, the various IMFs of
\citet{b:scalo86} would decrease the conversion factor to anywhere
from $1.7 \times 10^{40}$ to $8.4 \times 10^{40}$
\citep{b:kennicutt94}, and that of \citet{b:kroupa01} would give a
conversion of $1.9 \times 10^{41}$ \citep{b:brinchmann04}.  However,
all \Halpha\ surveys suffer from this bias equally, and so as with
$H_{0}$, we convert other \Halpha\ surveys to our chosen IMF before
comparison.

\subsection{Flux corrections}

To calculate star formation rates, we first have to correct the flux
data for foreground and internal extinction, [\NII] contamination, and
stellar absorption.  The foreground extinction corrections are
accomplished using the \citet{b:schlegel98} extinction maps.  Since our
selection avoids the Galactic plane and the Magellanic clouds, the
foreground extinction correction increases the derived values of
\lha\ and $l_R$ by an average of only 10\%.  Stellar absorption,
as explained in Paper I, causes our measurements to consistently
underestimate \Halpha\ flux by 2 -- 6\% \citep{b:brinchmann04}.  To
adjust for this, we increase each of our \Halpha\ fluxes by 4\%.

For the [\NII] correction and internal dust absorption,
$A(\Halpha)_{\rm int}$, we adopt the relationships with $R$ band
absolute magnitude prior to dust absorption corrections, $M_R'$, given
by \citet{b:helmboldt04} and converted to the AB magnitude system, as
explained in Paper I.  Each galaxy's individual [\NII]/\Halpha\
correction is based on its \HI\ velocity, velocity width, and the
narrow-band transmission profile, as detailed in Paper I.  Integrated
over our entire sample, the [\NII] correction decreases \lha\ by 15\%;
$l_R$ is not affected.  The internal extinction correction
$A(\Halpha)_{\rm int}$ increases our estimate of \lha\ by a factor of
2.1 (0.82 mag).  As noted in Paper I, we assume that the $R$ band
internal dust absorption, $A(R)_{\rm int}$, is half that of
$A(\Halpha)_{\rm int}$, due to the well-known phenomenon of increased
extinction in \HII\ regions compared to the field
\citep{b:fanelli88,b:calzetti94}, and so $l_R$ increases by 0.41
magnitudes after extinction correction.

\section{Results} \label{s:results}

\subsection{Volume densities} \label{s:voldens}

The $R$ band luminosity density, corrected only for Galactic
extinction, is found to be
\begin{equation}
l_R' = (4.4 \pm 0.7) \times 10^{37}\ h_{70}\ \rm erg\ s^{-1}\
\ang^{-1}\ Mpc^{-3}
\end{equation}
while the \Halpha\ luminosity density, corrected for [\NII]
contamination and Galactic extinction, is
\begin{equation}
\lhap = (9.4 \pm 1.8) \times 10^{38}\ h_{70}\ \rm erg\ s^{-1}\ Mpc^{-3}.
\end{equation}
With our adopted IMF and resultant star formation rate conversion,
Eq.~\ref{e:sfr}, the local star formation rate density (uncorrected for
internal extinction) is
\begin{eqnarray}
\log(\SFRDp\ [\Msun\ \mbox{\rm yr}^{-1}\ \mbox{\rm Mpc}^{-3}]) &=&
-2.13\ ^{+0.08}_{-0.09}(ran.) \\ &\pm& 0.03(sys.) + \log(h_{70})
\nonumber
\end{eqnarray}

Corrected for internal dust extinction, these equations become:
\begin{eqnarray}
l_R\ &=& 7.0\ ^{+1.5}_{-0.3} \times 10^{37}\ h_{70}\ \rm erg\ s^{-1}\ \ang^{-1}\ Mpc^{-3} \\
\lha &=& 2.0\ ^{+0.6}_{-0.4} \times 10^{39}\ h_{70}\ \rm erg\ s^{-1}\ Mpc^{-3}
\end{eqnarray}
\begin{eqnarray}
\log(\SFRD\ [\Msun\ \mbox{\rm yr}^{-1} \mbox{\rm Mpc}^{-3}]) &=&
-1.80\ ^{+0.13}_{-0.07}(ran.) \\ &\pm& 0.03(sys.) + \log(h_{70})
\nonumber
\end{eqnarray}
The uncertainties in each variable are explained in detail in
subsections~\ref{s:random} and \ref{s:system}.

\begin{figure}
\centerline{\hbox{\includegraphics[width=10.5cm]{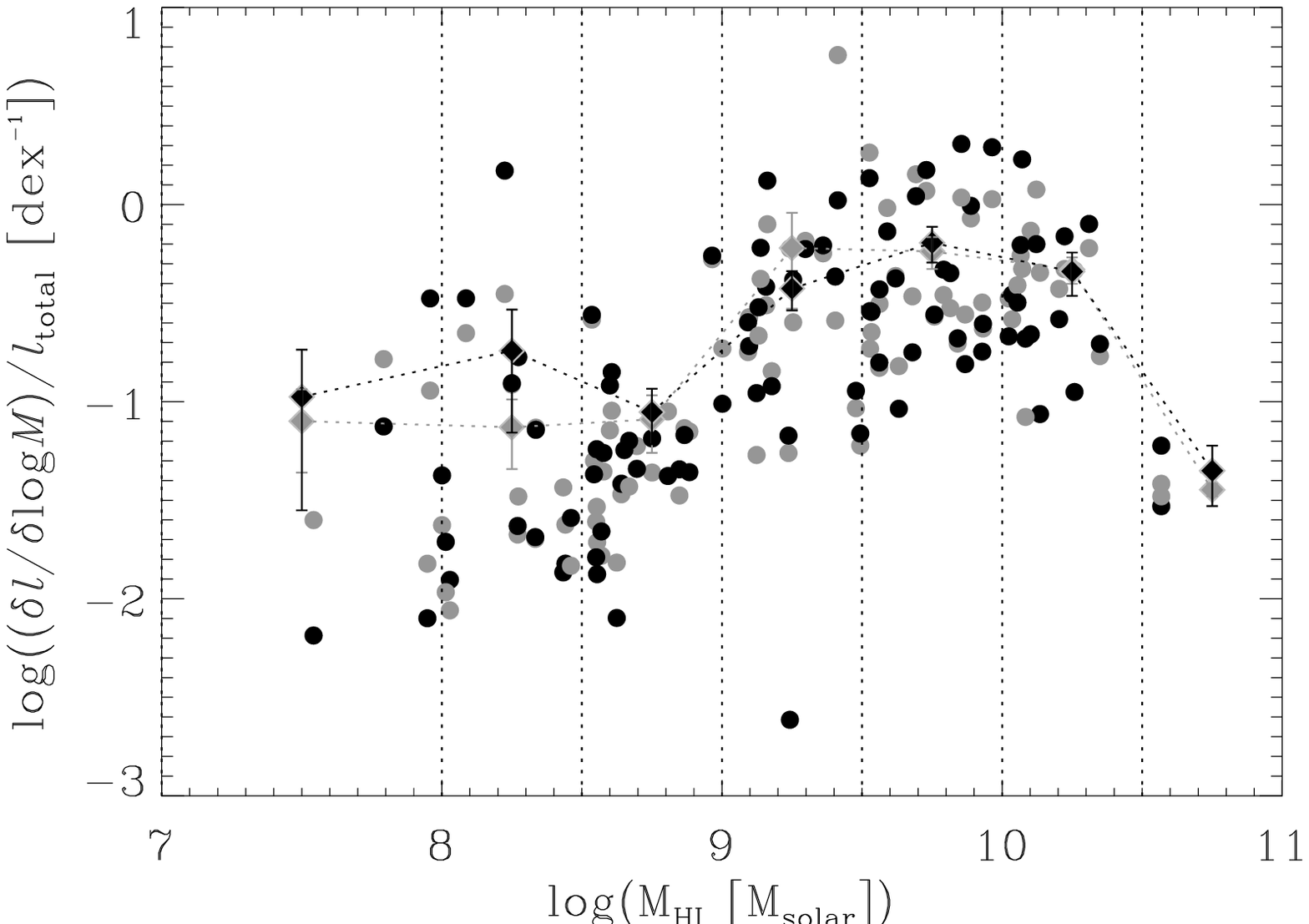}}}
\caption[]{Fraction of total luminosity density per \HI\ mass decade.
Dark symbols represent \Halpha , while lighter symbols denote $R$ band
luminosities.  Circles represent the value for the individual SINGG
galaxies.  Diamonds and error bars are the average values and standard
deviations of mean for each mass bin.  All values are corrected for
internal extinction.
\label{f:ldensity}}
\end{figure}

Additionally, we derive the \HI\ mass density to be
\begin{equation}
\rhoHI = 5.17 \pm 0.38 \times 10^{7}\ h_{70}\ \Msun\ \rm Mpc^{-3}
\end{equation}
and the mean number density of \HI -rich galaxies to be
\begin{equation}
n = 0.112\ ^{+0.017}_{-0.024}\ h_{70}\ \rm Mpc^{-3}
\end{equation}
within our mass range, $7.0 \le \log(\MHI/\Msun ) \le 11.0$.  Unlike
the luminosity densities, this number density is completely dependent
on the lower boundary chosen, as the integrated function increases as
\MHI\ decreases.  Fig.~\ref{f:ldensity} shows the dependence of the
luminosity densities (and, by extension, derived quantities such as
\SFRD ) on \MHI , along with the systematic uncertainty in the total
value due to each bin.  The plotted quantity is the fraction of the
total luminosity density coming from one decade of mass.

For comparison, we also derive \SFRD\ and \rhoHI\ for other mass
functions, using the simpler distance model favored by \HIPASS ; these
values are given in Table~\ref{t:himf}.  Only the \HIMF\ weighting of
each \Halpha\ source is altered; the distances and fluxes remain
unchanged.  While the differences in individual \HIMF\ parameters are
relatively small, the resulting integrated mass and luminosity
densities in Table~\ref{t:himf} vary by 0.20 -- 0.22 dex, primarily
due to the low-mass slope, $\alpha$.  While this discrepancy is
comparable to the total uncertainties in each density, it is
significantly larger than the $\sim 0.03$ dex uncertainty caused by
the uncertainties in the \HIMF\ parameters themselves.

\subsection{Random errors} \label{s:random}

Since the calculations of luminosity densities involve many different
variables, most of which have their own uncertainties, the simple
methods of error propagation would result in heavily correlated
uncertainties.  To better quantify the random uncertainties, we
utilize Monte Carlo and ``bootstrap'' algorithms.

We consider the following sources of random error: (1) the \HIMF: the
limited number of \HIPASS\ sources used to derive each \HIMF\ results
in uncertainties in the parameters used to fit that \HIMF, and hence
on luminosity densities; (2) error due to the limited SINGG SR1
sample; (3) $R$ and \Halpha\ flux uncertainties due to sky
subtraction; (4) \Halpha\ flux uncertainty due to continuum
subtraction; (5) the uncertainty in the \Halpha\ flux calibration, and
(6-7) error due to the dispersion in the fits used to generate our
[\NII] and $A(\Halpha)_{\rm int}$ corrections.  The random errors are
presented in Table~\ref{t:error}; we consider each of these terms in
detail.

To estimate the uncertainty due to the \HIMF , we create one hundred
realizations of the \HIMF\ using a bootstrap resampling of the
original data in \HICAT\ and adopting the \citet{b:mould00} distance
model.  The \HIMF\ in each realization is created using the same two
dimensional stepwise maximum likelihood technique of
\citet{b:zwaan05}.  Each realized \HIMF\ is fit to a Schechter
function, which is used in Eq.~\ref{e:rhox2} to determine the
resulting luminosity densities of each realization.  The random
uncertainty in each luminosity density due to the \HIMF\ is then the
dispersion about the mean luminosity density for all the realizations.

To quantify the SR1 sampling error we use a ``bootstrap'' resampling
method, drawing 92 objects at random (with duplication allowed) from
our data set.  As with our other uncertainties, this randomization is
repeated ten thousand times, with the overall sampling uncertainty
defined as the standard deviation of the resulting distribution.

The errors due to the uncertainties in the continuum scaling ratio and
background sky level for each galaxy are quantified through another
form of Monte Carlo logic.  For each of ten thousand iterations, we
vary the sky level or continuum ratio within Gaussian distributions
having the uncertainties derived from our error models, which in turn
alter the measured flux (and by extension, the star formation rate)
for each galaxy.  Again, the luminosity densities are recalculated for
each iteration, and the dispersion about the mean is quoted as the
resulting uncertainty.  Additionally, we have an uncertainty due to
our flux calibration method, as explained in Paper I; we estimate this
uncertainty to be 0.04 magnitudes for images requiring our 6568/28
narrow-band filter, and 0.02 magnitudes for all others.

To find the random error due to our $A(\Halpha)_{\rm int}$ and [\NII]
corrections, we use the $M_R'$ fits given in \citet{b:helmboldt04},
each of which has a dispersion of 0.23 dex (Helmboldt, priv.\ comm.).
We make a series of realizations of our sample; within each
realization, each galaxy's $\log(A(\Halpha)_{\rm int})$ or
$\log(F_{\rm [NII]6583}/F_{\rm H\alpha})$ correction is perturbed by a
Gaussian random deviate with the above dispersion.  Luminosity
densities are re-derived, and the dispersion about the mean simulated
luminosity density is our random error.  Separate sets of realizations
are done to determine the errors due to $A(\Halpha)_{\rm int}$ and
those due to the [\NII] correction, with all other terms held fixed.

Applied to the SINGG sample, we find the uncertainties as listed in
Table~\ref{t:error}.  We find that the random uncertainties are
dominated by the internal dust extinction (for the corrected
luminosity densities only), the SR1 sampling error, and to a lesser
extent the \HIMF\ uncertainties.  While the sky and continuum
subtraction uncertainties dominate the measurements of many individual
galaxies, when evaluated over our entire sample their contributions to
the error budget are relatively small.

\subsection{Systematic errors} \label{s:system}

The SINGG results also suffer from a series of systematic
uncertainties.  We consider the following sources of systematic error:
(1 -- 2) uncertainty in the zeropoints of our [\NII] and $A(\Halpha)_{\rm
int}$ corrections; (3) variation due to our choice of distance model;
(4) variation due to our choice of \HIMF .

In addition to the dispersion mentioned in Section~\ref{s:random}, our
$A(\Halpha)_{\rm int}$ and [\NII] corrections include an uncertainty
in the fit from \citet{b:helmboldt04} itself.  This zeropoint
uncertainty corresponds to the discrepancy in luminosity density for
fits one standard deviation of mean away from the best fit, generated
from the 196 sources of \citet{b:jansen00}.

Our distance model is a significant source of systematic uncertainty,
as most SINGG targets are located within a distance of 20 Mpc.
Variations in the distances used can result in large changes in
observed luminosities as well as the underlying \HI\ Mass Function,
and as a result alter the derived star formation rates.  To quantify
this effect, we calculate \SFRD\ using both our default
\citet{b:mould00} distance model as well as the simpler local-group
model used by \citet{b:zwaan05}, and quote the difference as our
uncertainty.

Finally, we calculate our luminosity densities using a variety of \HI\
mass functions and the local-group distance model, with results shown
in Table~\ref{t:himf}.  While the \HIPASS\ team has produced several
\HI\ mass functions, each supersedes the one before, with the work of
\citet{b:zwaan05} comprising the most complete version.  As a result,
while the resulting values of \SFRD\ vary by non-negligible amounts,
the earlier \HIMF\ values are only for comparison purposes, and we do
not include this error in our final uncertainties.


Our quoted systematic uncertainties are completely dominated by our
choice of distance model.

\subsection{Equivalent width}

We define the volume-averaged \Halpha\ equivalent width to be the
ratio of \Halpha\ flux density to $R$ band flux density.  We find the
volume-averaged equivalent width of our sample to be
\begin{equation}
EW'(\Halpha) = 21.2\ ^{+4.2}_{-3.5}\ \ang 
\end{equation}
without internal extinction corrections, and 
\begin{equation}
EW(\Halpha) = 28.8\ ^{+7.2}_{-4.7}\ \ang 
\end{equation}
after all corrections have been applied.  The difference between these
two values is due to the differential nature (line versus continuum)
of the extinction law applied \citep[][Paper I]{b:calzetti94}.

The \Halpha\ equivalent width is a measure of the rate of
transformation of the ISM into stars, relative to the existing stellar
population.  Comparison of the dust-corrected equivalent width to the
theoretical curves of \citet{b:kennicutt94} yields the birthrate
parameter $b$, corresponding to the ratio of the star formation rate
at the present time to that averaged over the age of the galactic
disk.  This interpretation is heavily dependent on the IMF used; for
our adopted \citet{b:salpeter55} IMF, we find
\begin{equation}
b = 0.26\ ^{+0.10}_{-0.06}
\end{equation}
which supports the common assertion that the cosmic star formation
density has decreased substantially since $z \sim 1.5$ within
star-forming galaxies.  Other IMFs result in greater values of $b$,
ranging as high as 2.8 for bottom-heavy IMFs such as
\citet{b:scalo86}.

\subsection{Gas cycling timescale} \label{s:tgas}

Using the SINGG data, we determine the gas cycling timescale
\tgas , the time it would take for star formation to process the
existing neutral and molecular ISM of a galaxy at its current rate of
star formation.  This value can be derived from \HI\ data if two
assumptions are made.  First, we assume that \HI\ emission occurs
predominantly at low optical depth.  Second, we assume that the ratio
of molecular hydrogen to neutral atomic hydrogen remains constant,
independent of the other characteristics of the galaxy.

To calculate \tgas\ we use the equation given in Paper I,
\begin{equation}
\tgas\ \approx 2.3 \left({\MHI\ \over SFR}\right)
\label{e:tgas}
\end{equation}
where ${\cal M}_{gas}$ = 2.3 \MHI, derived from the observations of
\citet{b:young96}, accounts for the typical molecular hydrogen as well
as the helium content of the ISM.

\begin{figure}
\centerline{\hbox{\includegraphics[width=10.5cm]{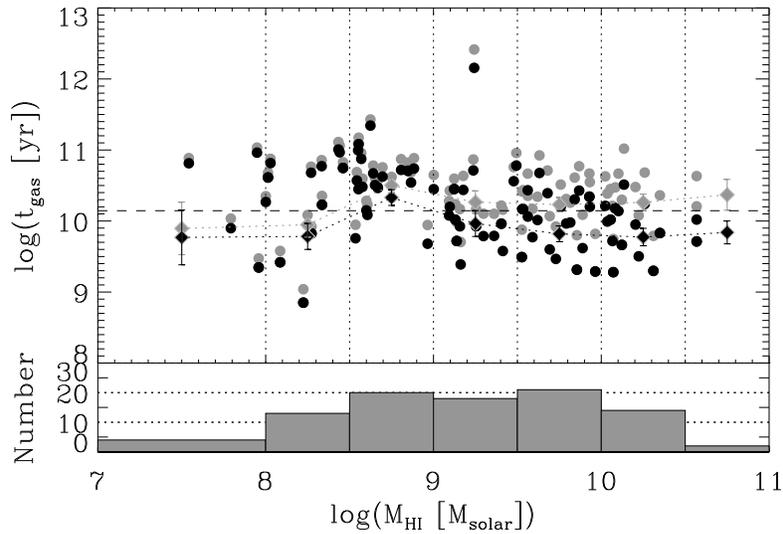}}}
\caption[]{Gas cycling timescale and histogram.  Circles are values
for individual galaxies.  Diamonds and error bars are the average
values and standard deviations of mean for each mass bin.  The dashed
line corresponds to the Hubble time (13.6 Gyr).  Dark symbols are
corrected for internal extinction, lighter symbols are uncorrected
values for the same galaxies.
\label{f:gastime}}
\end{figure}

The values of \tgas\ for our array of mass bins are given in Table
\ref{t:bindata}.  The volume-averaged gas cycling timescale for our
sample can be derived from \tgas\ $\approx 2.3 \left(\rhoHI
/\SFRD \right)$; with the values for \rhoHI\ and \SFRD\ presented in
Section~\ref{s:voldens}, we find this timescale to be $\tgas\ = 7.5\
^{+1.3}_{-2.1}$ Gyr.  This is somewhat less than the Hubble time,
consistent with the previous findings of \citet{b:kennicutt94}.
Fig.~\ref{f:gastime} shows the gas cycling timescales for each of the
SINGG galaxies, as well as the average within each \HI\ mass bin.
While the galaxy-to-galaxy variation in \tgas\ is substantial (0.54
dex), the mean value does not appear to vary greatly with \MHI .

Our choices of distance model and \HI\ Mass Function have much smaller
impacts on the cosmic \tgas\ than they do on \SFRD, and have no effect
on the gas cycling times of the individual galaxies within our sample.
Using the other \HI\ Mass Functions given in Table~\ref{t:himf} result
in cosmic \tgas\ values of 7.0 to 7.3 Gyr, well within the error bars
of our adopted \HIMF .

One should not take these gas cycling timescales too literally, as
several of the assumptions behind them may be questioned.  Much of the
neutral interstellar medium may have a large optical depth, which
would imply larger values of \MHI\ than measured, causing \tgas\ to be
somewhat underestimated.  This effect probably results in a
discrepancy of less than 20\% \citep{b:haynes84}.  As mentioned in
Paper I, the observed ratio of $\left(\MHtwo /\MHI \right)$ used to
derive Eq.~\ref{e:tgas} has a dispersion of 0.58 dex, or almost a
factor of 4; the ratio of CO luminosity to H$_{2}$ mass also has a
large uncertainty.  Hence, the \tgas\ estimate of any single galaxy is
likely to be highly uncertain.  Similarly, we have not accounted for
systematic effects.  It is well known that more massive and higher
surface brightness galaxies are easier to detect in CO emission than
dwarfs.  In addition, our scenario does not include the ``hot phase''
of the ISM.  This accounts for the X-ray emitting halos around
galaxies as well as the intracluster medium and the intergalactic
medium.  This is likely to be the largest baryonic component of the
universe \citep{b:fukugita98}.  We also do not account for the return
to the ISM of material processed by stars, the stellar yield.  Because
of these limitations we are careful not to imply that \tgas\ is a
consumption timescale; rather, it is simply the time it would take for
the present rate of star formation to form the mass of observed
neutral ISM and inferred molecular ISM into new stars.


\subsection{Dynamical parameters} \label{s:dyn}

The dynamical mass located within the optical radius of a galaxy,
\Mdyn, is a useful way to estimate its mass (including both
baryonic and dark matter) from easily observable quantities.  The
corresponding orbital time, \torb , is also used for comparison with
various models of galaxy evolution.  To find these values, we first
approximate the circular velocity of the edge of the galaxy from the
FWHM spread in the velocities measured by \HIPASS , corrected for the
estimated inclination of each galaxy.  That is,
\begin{equation}
v_{\rm circ} = \left(\frac{W_{50}}{2\ sin(i)}\right).
\label{e:vcirc}
\end{equation}
where our inclination angle, $i$, is defined as 
\begin{equation}
sin^{2}(i) = \left(\frac{a^2 - b^2}{a^2 - c^{2}}\right)
\end{equation}
\citep{b:bottinelli83}, where $a,c$ are the axial lengths of the
oblate spheroid fit to each galaxy, and $a,b$ are the axial lengths of
the elliptical aperture used to measure optical fluxes.  Since we do
not wish to include any dependence on morphological classifications in
the SINGG methodology, we assume a constant $(c/a) = 0.20$, except in
three cases where our estimated $(b/a)$ is less than this amount,
where we simply set $c = b$.  To reduce uncertainty in \Mdyn\ we
exclude the 13 \HI\ targets with multiple emission line galaxies, as
well as 14 additional face-on galaxies ($(b/a) \geq 0.8$, or $i \leq
38^{\circ}$).

We then find the mass contained within the observed limit of each
galaxy to be
\begin{equation}
\Mdyn = \left(\frac{v_{\rm circ}^{2}\ \rmax}{G}\right)
\end{equation}
and the orbital time to be
\begin{equation}
\torb = \left(\frac{2\pi\ \rmax}{v_{\rm circ}}\right).
\label{e:torb}
\end{equation}

\begin{figure}
\centerline{\hbox{\includegraphics[width=16.0cm]{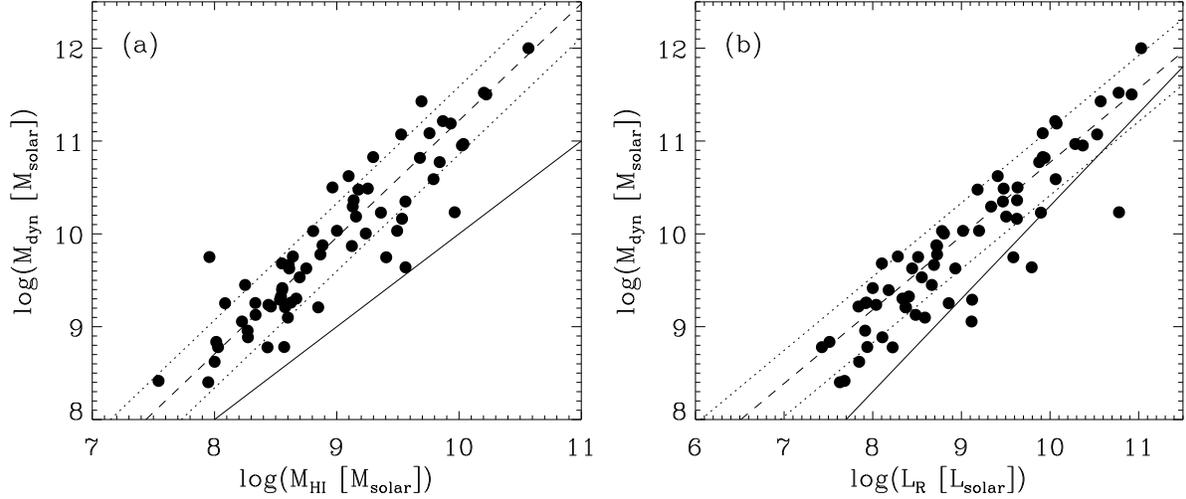}}}
\caption[]{Dynamical mass as a function of (a) \HI\ mass and (b) $R$
band luminosity.  The dashed lines show the best fits, with dotted
lines showing the dispersion for each.  Solid lines show (a) \Mdyn\ =
\MHI\ and (b) \Mdyn\ = $2\ L_R$ in solar units (where M/$L_{R}
\approx 2$ is typical of gas-rich galaxies).
\label{f:massscatter}}
\end{figure}

\Mdyn\ is evaluated using the radius at which we truncate our light
profiles, \rmax . This radius is found using only the optical light of
the galaxy, therefore matter located outside it will not be included.
\Mdyn\ is not a true total mass; rather, it is a crude estimate of the
mass contained within the optical radius.  Since \HI\ typically
extends beyond \rmax\ with a flat rotation curve, the total mass
(including dark matter) is larger, and our \Mdyn\ estimates are thus
lower limits to the total mass.  \Mdyn\ is compared with our \HIPASS
-derived values of \MHI\ in Fig.~\ref{f:massscatter}a, and with our
observed $R$ band luminosity, $L_R$, in Fig.~\ref{f:massscatter}b.
After excluding galaxies as explained above and fitting with a robust
least absolute deviation fit, we find:
\begin{eqnarray}
\log(\Mdyn) =& 1.26\ \log(\MHI) &-\ 1.36 \pm 0.37 \label{e:mdynvsmhi}\\
\log(\Mdyn) =& 0.79\ \log(L_{R}) &+\ 2.83 \pm 0.36 \label{e:mdynvslr}
\end{eqnarray}
Eq.~\ref{e:mdynvsmhi} implies that the fraction of \Mdyn\ made of \HI\
is typically more than six times higher for our lowest \HI\ masses
(\MHI\ $\approx 10^{7.5}$ \Msun ) than it is for those at our
high-mass extreme (\MHI\ $\approx 10^{10.6}$ \Msun ).  Likewise, if we
were to assume that $\cal M$/$L$ remains constant over our range of
dynamical masses, Eq.~\ref{e:mdynvslr} would imply that the visible
mass fraction is up to 10 times larger for our most massive galaxies
as for our least massive.  For any further equations requiring \Mdyn ,
we use the fit of Eq.~\ref{e:mdynvslr} to set the dynamical mass of
those galaxies exluded from our fit because of inclination or multiple
\Halpha\ sources.  While Eq.~\ref{e:mdynvslr} is reminiscent of the
classic relations of \citet{b:tully77}, our \Mdyn\ depends both on the
width of the \HI\ velocity profile and the optical radius, \rmax.  As
a result, our result should not be directly compared to the
Tully-Fisher relationship, $\Mstar \sim v_{\rm circ}^{3.1}$.

We are primarily interested in \torb\ because it lets us address the
redshift evolution of SFR in conjunction with the star formation law
of \citet{b:kennicutt98}, which states that the global star formation
law of galaxies is equivalent to galaxies converting 21\%\ of their
ISM mass within one orbital time \torb\ evaluated at the radius where
the \HII\ region distribution is truncated, \rHII .  We have not
measured \rHII\, as such, for our sample.  However, Paper I provides
two radii that should bracket this: $r_{90}(\Halpha)$, the radius
enclosing 90\%\ of the \Halpha\ flux, and \rmax; for most sources,
\rmax\ contains all the discernable emission in both $R$ and \Halpha .
We evaluate the orbital time, \torb , at these two radii.  As with our
determination of \Mdyn, \torb\ is ill-determined for galaxies which
have $i \leq 38^{\circ}$ or are in multiple ELG systems.  In those
cases we estimate $\log(\torb)$ from $L_{R}$ using least absolute
deviation fits, similar to Eq.~\ref{e:mdynvslr}:
\begin{eqnarray}
\log(\torb(\rmax)) &=& 0.099\ \log(L_{R}) - 1.04 \pm 0.20 \\
\log(\torb(r_{90}(\Halpha))) &=& 0.106\ \log(L_{R}) - 1.35 \pm 0.23
\end{eqnarray}
These fits are then used to set orbital times for the excluded
galaxies, in the same manner as for \Mdyn.

As with \HI\ mass, we use \Mdyn\ to find the local dynamical mass
density, $\rho_{\rm dyn}$, with Eq.~\ref{e:rhox2}.  We find
\begin{equation}
\rho_{\rm dyn} = 9.3\ ^{+1.4}_{-1.6} \times 10^8\ h_{70}\ \Msun\
\rm Mpc^{-3}.
\end{equation}
Again, we emphasize that this is only the mass density of the local
universe residing within the optical radii of \HI -rich galaxies.  Of
this density, 15\%\ is found in galaxies with multiple \Halpha\
sources, while an additional 34\%\ is found in the face-on galaxies
excluded from Eq.~\ref{e:mdynvslr}.  Comparing to our \HI\ mass
density, \rhoHI , we estimate that 5.6\%\ of the dynamical mass of
local galaxies consists of neutral hydrogen in stars.

\section{Discussion} \label{s:disc}

\subsection{Completeness} \label{s:complete}

One of our goals is to determine how representative an \HI -selected
sample is.  To do this, we compare the $R$ band luminosity density
$l_R'$ and dynamical mass density $\rho_{\rm dyn}$ derived from an \HI
-selected sample to those derived from more ``complete'' samples, such
as the full SDSS sample set.

Our uncorrected value for $l_R'$, $4.4 \pm 0.7 \times 10^{37}\ h_{70}$
erg s$^{-1}$ \ang$^{-1}$ Mpc$^{-3}$, compares well to the SDSS-derived
$^{0.1}r$ and $^{0.1}i$ band values of \citet{b:blanton03}, $6.17$ and
$6.70 \times 10^{37}\ h_{70}$ erg s$^{-1}$ \ang$^{-1}$ Mpc$^{-3}$
respectively.  Interpolating between these two bands by wavelength
gives $l_R' \approx 6.30 \times 10^{37}\ h_{70}$ erg s$^{-1}$
\ang$^{-1}$ Mpc$^{-3}$ for the Sloan survey, which we will use as our
``cosmic'' $R$ band density.  Our estimated $l_R'$ is approximately
70\% of this value; therefore, the majority of the stars comprising
the derived SDSS $R$ band luminosity density are located in galaxies
containing measurable quantities of \HI .  If we were to assume that
our $l'_{R}$ is low due to gas-poor galaxies (ellipticals) being
absent from our sample, while our \lhap\ is complete, then we can
derive a corrected ``cosmic'' $EW(\Halpha) \approx\ \lha/(l_{R}/0.70)
= 20.3\ ^{+5.9}_{-3.5}\ $\ang .  This reduces the Salpeter-derived
birthrate parameter to $b \approx 0.16\ ^{+0.07}_{-0.03}$.

Note that the SR1 sample size implies that certain rare types of
galaxies simply won't be represented in these results.  For instance,
ultraluminous IR galaxies \citep[ULIRGs, defined as $L_{IR} \geq
10^{12} \Lsun$,][]{b:sanders96} are rare enough that the volume
contained within the \HIPASS\ redshift and declination limits would
only include roughly one ULIRG.  As the sample used in this paper
includes only 92 out of the 4315 \HIPASS\ targets, it was extremely
unlikely that any ULIRGs would be included in our observations.

When our dynamical mass density $\rho_{\rm dyn}$ is expressed as a
fraction of the critical Einstein-de Sitter mass density ($\rho_{crit}
= 1.36 \times 10^{11}\ h^{2}_{70}\ \Msun\ $Mpc$^{-3}$), we find
$\Omega_{\rm dyn} = 0.0068\ h^{-1}_{70}$, or 2.3\% of $\Omega_{0}$ in
the concordance cosmology.  For comparison, \citet{b:cole01} estimates
the the $z \approx 0$ mass density of stars, $\Omega_{\rm stars}$, to
be between 0.0029 (from an IR luminosity function) and 0.0020 (from
star formation tracers).  Similarly, the \HI\ Mass Function of
\citet{b:zwaan05} gives the local \HI\ mass density to be $\approx
0.0004\ h^{-1}_{70}$, which would imply a gas density on the order of
$\Omega_{\rm gas} = 0.0009\ h^{-1}_{70}$ using the same logic as in
Section~\ref{s:tgas}.  As a result, our $\rho_{\rm dyn}$ implies that
around half of the mass located within the optical disks of nearby
galaxies consists of gas and stars, with the remainder most likely
consisting of dark matter.

This comparison must be viewed with some caution; our $\rho_{\rm
dyn}$ is a crude estimate of all matter within the optical radii of
\HI -selected galaxies, including substantial quantities of dark
matter.  However, it excludes galactic gas (especially \HI ) extending
beyond the optically-selected \rmax , and also samples little of the
hot plasma which resides in galaxy clusters and the intergalactic
medium.  We also do not sample any of the mass in \HI -free galaxies,
predominantly early-type galaxies.  This comparison only shows that
the mass we sample is comparable to the baryon content of the local
star-forming galaxies, and that the ratio of baryonic matter to dark
matter within these galaxies is substantially different than that of
the universe as a whole.

\subsection{\SFRD\ and \lha\ in context}

\begin{figure}
\centerline{\hbox{\includegraphics[width=16.0cm]{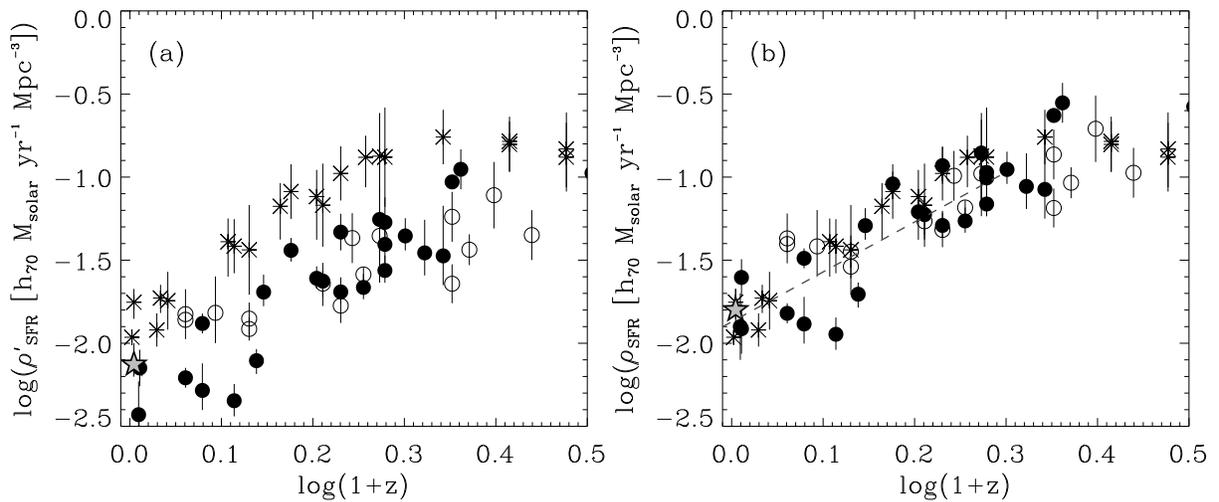}}}
\caption[]{Star formation rate density, (a) without and (b) with
corrections made for internal dust extinction.  Solid circles are
emission-line surveys (usually \Halpha ).  Hollow circles are UV
surveys.  Asterisks are IR/sub-mm surveys.  The star at $z \approx 0$
is the SINGG value.  Other values are drawn from \citet{b:hopkins04}
and \citet{b:perez05}.  The dashed line in (b) corresponds to the best
fit from $0 < z < 1$.
\label{f:madau}}
\end{figure}

To compare our value of \SFRD\ to those found by other studies, we
refer to \citet{b:hopkins04}, which compiled the results of 33 other
star formation rate density papers and corrected each to a uniform
$\Lambda$CDM cosmology, a \citet{b:salpeter55} IMF, and with a Hubble
constant of $H_{0}$ = 70\ km s$^{-1}$ Mpc$^{-1}$, $\Omega_{0}$ = 0.3,
and $\Omega_{\Lambda}$= 0.7.  We also include the IR-derived \SFRDz\
data of \citet{b:perez05}.  The results before internal extinction
correction are plotted in Fig.~\ref{f:madau}a, while the corrected
values are plotted in Fig.~\ref{f:madau}b.

In Paper I we assert that SINGG is inherently less biased than other
surveys, due to selecting sources by ISM content at radio wavelengths.
Objective-prism surveys limit their samples by equivalent width and
surface brightness, while an \HI -selected sample can include diffuse
sources or those with little star formation.  If this is the case, we
should expect to recover more star formation in the local universe.
However, there are many steps involved in turning measured fluxes into
luminosity density estimates; since techniques and assumptions vary
between groups it is important to compare our results in as consistent
a form as possible.  Few previous studies present actual \Halpha\
luminosity densities, and so our primary comparison will be between
values of \SFRDpz\ corrected to our adopted IMF and extrapolated to $z
= 0$.  While some studies have only provided extinction-corrected
\SFRDz\ estimates, in most cases we are able to work backwards using
published corrections to derive \SFRDpz\ for each.

When the surveys located at $0.0 \le z \le 1.0$ are linearly fit to a
simple \SFRDz\ = \SFRDo\ $(1+z)^{\beta}$ relationship, we find $\beta
= 3.00 \pm 0.13$, which matches well with the $\beta = 3.2\
^{+0.7}_{-0.2}$ of \citet{b:lefloch05} and the $\beta = 3.1 \pm 0.5$
of \citet{b:perez05}.  As a result, we will use this proportionality
to extrapolate the star formation rate densities of low-redshift
surveys to $z \sim 0$; the results are given in Table~\ref{t:sfrd}.

The relative positions of the data points in Fig.~\ref{f:madau}a,b
change due to the internal extinction correction.  While the spread in
points noticeably decreases between the two figures, we cannot claim
to measure the highest corrected \SFRDo .  This is primarily due to
our relatively mild extinction correction of 0.82 magnitudes, while
the other samples noted in Table~\ref{t:sfrd} have corrections ranging
from 1.0 to 1.4 mag.  Although mild, our correction correlates well
with unpublished work from the 2dF Galaxy Redshift Survey
\citep{b:folkes99}, which used the Balmer decrement to estimate the
\Halpha\ extinction of 160,000 line emitters; the typical value was
found to be approximately 0.8 magnitudes.  This discrepancy in
$A_{int}(\Halpha)$ is most likely caused by the same luminosity bias
mentioned above.


Even with a value of \SFRD\ comparable to other surveys, it is too
early to declare a consensus because of the differences between the
samples.  For instance, we assert that the objective-prism selection
method results in a large, consistent bias against non-starburst
galaxies.  Here we define starburst galaxies as those with \Halpha\
equivalent widths within their half-light radii of $EW'_{50}(\Halpha)
\geq 50 $\ang , without correction for internal extinction.  According
to \citet{b:heckman98}, starbursts are estimated to comprise 15 --
20\% of the population of the local universe.  In the UCM
spectroscopic survey, 72\% of the sample consists of galaxies with
equivalent widths above this threshold.  For SINGG SR1, 16 out of the
110 \Halpha\ sources (14.5\%) met this criterion; these sources are
collectively responsible for 25\%\ of our final star formation rate
density.  If we assume that the \HI -selected SINGG sample set is not
significantly biased towards or against these galaxies, emission
line-selected surveys (such as UCM) should be significantly
underestimating the value of \SFRD\ due to underrepresentation of the
non-starburst galaxies which appear to produce the majority of the
star formation in the local universe.

\subsection{Breakdown of luminosity density and \SFRD} \label{s:cumuhists}

\begin{figure}
\centerline{\hbox{\includegraphics[width=16.0cm]{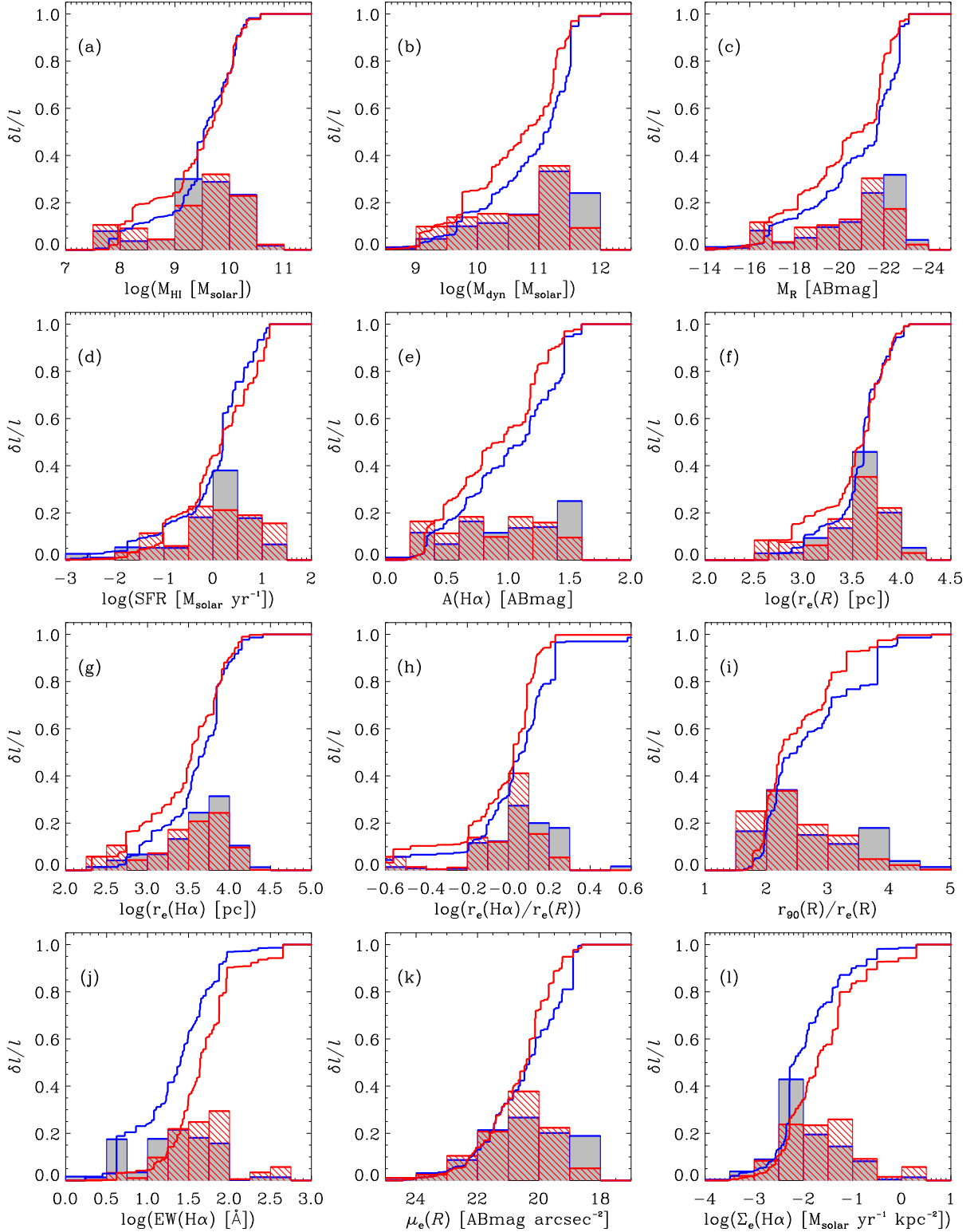}}}
\caption[]{Fraction of the total luminosity density, $l$, as a
function of various quantities.  Red lines correspond to \Halpha\
luminosity, while blue lines correspond to $R$ band luminosity.
Cumulative values as well as binned histograms are given for each.
\label{f:moreplots}}
\end{figure}

Fig.~\ref{f:moreplots} shows the observed contributions to the $R$
band and \Halpha\ luminosity densities as a function of a variety of
different quantities; this allows a comparison between our sample and
other local samples of galaxies.  Table~\ref{t:morestat} gives the
values of each parameter at the 10$^{th}$, 25$^{th}$, 50$^{th}$,
75$^{th}$, and 90$^{th}$ percentiles of each luminosity density.  In
all cases, extinction and [\NII] corrections have been applied as
appropriate.

Plot (a) relates the luminosity densities to the \HIPASS\ \HI\ mass.
We find that 72\% of $l_{R}$ and 70\% of \lha\ are found in galaxies
with \HI\ masses below our value of \Mstar(\HI) .  However, 77\% of
\rhoHI\ falls below this value, so the stellar luminosity density and
\SFRD\ are slightly weighted towards higher \HI\ masses than \rhoHI .

Plot (b) relates the luminosity densities to the dynamical mass, with
the fit from Fig.~\ref{f:massscatter}b used to estimate dynamical
masses for those galaxies with multiple sources or low axial ratios.
Galaxies with low dynamical masses contribute substantially more to
the \Halpha\ luminosity than the $R$ band density, and the SINGG
sample extends across a wide range of \Mdyn\ values.

Plot (c) relates $l$ to the $R$ band luminosity by way of the absolute
magnitude, M$_{R}$.  As expected, galaxies with high $R$ band
luminosities contribute more to $l_R$ than to \lha.  For comparison,
the study of \citet{b:brinchmann04} used a stellar mass function with
its knee at $\Mstar = 10^{10.95} \Msun$; as this is a stellar mass
(not simply \HI\ mass), it should be approximately proportional to the
$R$ band luminosity.  The 10$^{th}$ and 50$^{th}$ percentiles for
\lha\ in \citet{b:brinchmann04} were quoted as $\log(\Mstar) = 9.0$
and $10.3$, respectively, a difference of 1.3 dex.  For comparison,
the corresponding SINGG values of M$_{R}$ for those percentiles differ
by 4.1 magnitudes (1.7 dex), as shown in Table~\ref{t:morestat}.

Plot (d) relates $l$ to the star formation rate of each galaxy
(directly proportional to \Halpha\ luminosity, as shown in
Eq.~\ref{e:sfr}.)  As expected, galaxies with high \Halpha\
luminosities contribute more to \lha\ than $l_R$.  However, the
fraction of \Halpha\ luminosity density caused by luminous galaxies in
SINGG is significantly lower than in other surveys.  Only 12.2\% of
the SINGG \lha\ is caused by galaxies with observed star formation
rates greater than 10.0 \Msun\ yr$^{-1}$ once extinction corrections
have been applied.  For comparison, we integrate the luminosity
functions quoted by other surveys; the UCM study of
\citet{b:gallego95} has 26.5\% of its \lha\ come from galaxies above
this threshold, while \citet{b:tresse98} has 23.1\%.  At the extreme
cases, \citet{b:sullivan00} has only 9.3\%, while \citet{b:perez03}
has 51.4\% of its luminosity density come from galaxies with star
formation rates above 10.0 \Msun\ yr$^{-1}$.  This further illustrates
that the apparent consensus in \SFRD\ may hide important differences
in the various surveys of local star formation.

Plot (e) relates $l$ to the narrow-band internal dust extinction
correction, A(\Halpha).  While the extinction correction in our
overall \Halpha\ luminosity density is 0.82 magnitudes, only 37\% of
the $R$ density comes from galaxies with smaller values, compared to
47\% of the \Halpha\ density.  That is, $l_{R}$ is weighted towards
the more luminous galaxies, which have higher internal extinctions.

Plots (f) and (g) relate $l$ to the effective (50\%\ flux) radii found
in the $R$ and \Halpha\ images, while plot (h) relates $l$ to the
ratio of these two radii.  Again, these plots support the
``downsizing'' model of star formation; galaxies with large radii tend
to contribute more to the $R$ band luminosity density than the
\Halpha\ density, while galaxies with small radii contribute
substantially more to \lha\ than $l_{R}$.  The luminosity density
contributions are evenly distributed over our range of
($r_{e}(\Halpha)/r_{e}(R)$).  Galaxies with centrally-located star
formation ($r_{e}(\Halpha)/r_{e}(R) \le 1.0$, suggestive of
starbursts) do not dominate the luminosity densities; this contradicts
what has been observed in other studies, such as the UCM survey of
\citet{b:gallego95}.

In all of the above plots, the smaller, less massive galaxies
contribute substantially more to the SINGG \Halpha\ luminosity density
(and by extension, \SFRD ) than they do to the $R$ band density.  This
reinforces the ``downsizing'' model, where star formation activity
shifts to smaller galaxies over time \citep{b:cowie96}.

Plot (i) relates $l$ to the ratio of the 90\% flux radius to the
half-light radius in the $R$ band.  The inverse of this ratio
(referred to as the ``concentration index'') has been examined by
\citet{b:shimasaku01} using SDSS data.  In that work, it was estimated
that the boundary between bulge-dominated ``early-type'' and
disk-dominated ``late-type'' galaxies occurred at a ratio of
$r_{90}(r')/r_{e}(r') = 3.03$.  Given that definition, we estimate
that 71\% of our $R$ density ((3.2\ $\pm\ 0.5) \times 10^{37} h_{70}$\
erg s$^{-1}$\ \ang$^{-1}$\ Mpc$^{-3}$ before extinction correction)
and 82\% of our \Halpha\ density can be attributed to ``late-type''
galaxies, assuming the $r$-derived boundary does not change when
shifted to our $R$ band.  For comparison, \citet{b:hogg02} find that
38\% of the SDSS $^{0.1}i$ band luminosity density comes from
``red-type'' (assumed to be bulge-dominated early-type) galaxies,
meaning that a total of $4.1 \times 10^{37}\ h_{70}$ erg s$^{-1}\
$\ang$^{-1}$\ Mpc$^{-3}$ can be attributed to late-type galaxies.  As
their definition of red galaxies was conservative, this percentage
should include some early-type galaxies.  Likewise, \citet{b:baldry04}
fit a bimodal distribution to observed galaxy colors to determine that
58\% of the SDSS $^{0.1}r$ band luminosity density comes from blue
galaxies (for a density of $3.6 \times 10^{37}\ h_{70}$ erg s$^{-1}$
\ang$^{-1}$\ Mpc$^{-3}$).  As a result, while the SINGG survey recovers
around 70\% of the SDSS $l_{R}$, we recover a larger fraction
(80 -- 90\%) of the luminosity density from late-type galaxies.

Plot (j) shows the dependence of $l$ on \Halpha\ equivalent width.
The \SFRD\ derived from the UCM spectroscopic survey
\citep{b:gallego95} uses only galaxies with equivalent widths larger
than 10 \ang .  According to our data, 24.1\% of the local
extinction-corrected $R$ band luminosity density and 4.5\% of the
corrected \Halpha\ luminosity density come from galaxies with
equivalent widths below this threshold.

Plots (k) and (l) relate $l$ to surface brightnesses, $\mu_{e}(R)$ and
$\Sigma_{e}$(\Halpha).  As expected, galaxies with high \Halpha\
surface brightnesses (i.e., starbursts) contribute a much larger
fraction of \lha\ than of $l_R$.  For comparison, the SDSS-derived
sample of \citet{b:blanton05} has a surface brightness limit of
$\mu_{e}(R)$ = 24.0 ABmag arcsec$^{-2}$; 1.1\%\ of our $l_R$ and
1.2\%\ of \lha\ come from galaxies below this cutoff.

\subsection{Recent redshift evolution in \SFRDz} \label{s:evolution}

Fig.~\ref{f:madau}b shows that the \SFRDz\ has declined by
approximately a factor of ten from $z \sim 1$ to the present.
Specifically, fitting the internal dust corrected data with $z \leq 1$
with a robust (outlier resistant) linear fit
\begin{equation}
\log(\SFRDz) = \log(\SFRD) + \eta z
\label{e:slope}
\end{equation}
yields $\log(\SFRD\ [\Msun\ $yr$^{-1}\ $Mpc$^{-3}]) = -1.73$ and $\eta =
1.02$ dex per redshift.  We can estimate the systematic uncertainties
in this fit by categorizing the \SFRDz\ data by the star formation
tracer used in each measurement: (1) optical emission lines; (2)
ultraviolet continuum; and (3) FIR or radio continuum.  This yields
$\log(\SFRD) = -1.80,\ -1.53,\ -1.94$, and $\eta = 1.05,\ 0.43,\
1.37$\ for survey types 1, 2, and 3, respectively.  This suggests that
the uncertainties in the zeropoint and slope are 0.20 and 0.47 dex,
respectively.

What causes the decrease in \SFRDz\ with cosmic time?  Recently,
\citet{b:kauffmann04} presented a model of the recent star formation
history of galaxies in the local universe that provides a useful
context to answering this question.  In their model, star formation
events occur when galaxies merge; subsequently, the SFR inside the
galaxy decays according to a prescription similar in nature to the
star formation laws of \citet{b:kennicutt98} and \citet{b:dopita94}.
They use a high-resolution Cold Dark Matter (CDM) simulation, identify
galaxy mergers with CDM halo mergers, and parameterize the form of
their star formation law by the average stellar mass density.
Essentially, the physics of their model can be separated into CDM
effects (the merger of halos), and baryonic physics (the star
formation law).  We can then rephrase the question above: is the
decrease in \SFRDz\ with cosmic time driven by CDM or baryonic
physics?  Are we seeing a recent decrease in the merger rate
of halos?  Or, are we seeing the secular decay of the SFR after a much
earlier decrease in the halo merger rate?

The dynamical information on our sample provided by the \HI\ line
widths provides a means to address this question.  The orbital time,
\torb , has been evaluated at two radii, $r_{90}(\Halpha)$ and \rmax ,
as explained in Section~\ref{s:dyn}.  Using the same methods as those
in Section~\ref{s:cumuhists}, we find that the 25\%, median, and 75\%\
contributions to \lha\ (and, by extension, \SFRD ) in our sample
occur for \torb\ = 0.30, 0.57, 0.60 Gyr when evaluated at
$r_{90}(\Halpha)$ and 0.56, 1.00, 1.03 when evaluated at \rmax .

We can use the orbital times to estimate the change in SFR with
redshift for the galaxies in our sample obeying Eq.~\ref{e:slope}.  If
we assume that the star formation law remains invariant over this
timescale, then consumption of 21\%\ of the ISM is equivalent to a
21\%\ decrease in the SFR over \torb\ \citep{b:kennicutt98}.
Translating \torb\ to a look back time, the logarithmic change in SFR
is given by
\begin{equation}
\frac{\delta\log(\rm SFR [\Msun\ yr^{-1}])}{\delta z} \approx
\frac{1.43\ h_{70}}{\torb\ [\rm Gyr]}
\end{equation}
in the same units as the slope $\eta$ in Eq.~\ref{e:slope}.  The
median \torb\ translates to $\delta\log(\rm SFR)/\delta z = $ 2.5, 1.4
as defined by $r_{90}(\Halpha)$ and \rmax , while the interquartile
ranges are 4.8 -- 2.4 and 2.6 -- 1.4, respectively.  We see that the
SFR of the galaxies in our sample will decay over a wide range of
timescales.  However, a galaxy with \torb\ similar to the median
galaxy in our sample will have its star formation decay at a rate
similar to that observed in the universe as a whole.

\section{Conclusions} \label{s:concl}

Our measurement of the local star formation rate density, $\log(\SFRD\
[\Msun\ \mbox{\rm yr}^{-1}\ \mbox{\rm Mpc}^{-3}]) = -1.80\ +
\log(h_{70})$, is similar to those of most previous $z \sim 0$
studies, as shown in Table~\ref{t:sfrd}.  When compared to the results
of other surveys, this agrees with the consensus that the star
formation rate density has decreased by an order of magnitude from a
redshift of $z \sim 1$ to the present.  While a consensus in \SFRD\
appears to be emerging, this may be largely illusory.  There remain
significant differences between the various surveys in terms of
extinction corrections and the contributions of various types of
galaxies to the totals.  Systematic effects are large and need to be
accounted for before a true consensus can emerge.

The decreasing \SFRDz\ with cosmic time has important implications for
models of galaxy evolution, as this trend suggests a shift from the
faster ``burst'' star formation process to the slower ``quiescent''
process, with the greater fraction of the total star formation
occurring in non-starburst galaxies.  This is supported by the results
presented in Section~\ref{s:evolution}, which suggest that the
decrease in \SFRDz\ with cosmic time is largely driven by the secular
decay in SFR after earlier accretion events.  That is, the current
evolution in \SFRDz\ is largely driven by interior baryon physics
rather than the merger of CDM dominated halos.  This interpretation is
consistent with the relatively small fraction of multiple ELG systems
and targets that look like recent mergers (15 out of 93 \HIPASS\
pointings included in SR1, see Paper I), as well as the work of
\citet{b:bell05}.  In contrast, at $z > 1$ field galaxies tend to have
a disturbed morphology suggesting a higher fraction of interacting and
merging systems and very few regular disk galaxies can be discerned.
It is tempting to speculate that $z \sim 1$ represents the epoch where
cosmic evolution transitions from being driven largely by CDM
interactions to being driven by the internal self-regulation of star
formation.



\acknowledgments

Partial finantial support for this work has been provided by NASA
grant NAG5-13083 (LTSA program) to G.R.\ Meurer, as well as the
Director's Discretionary Research Fund of the Space Telescope Science
Institute.  We would like to thank the many people whose feedback
helped in the development of this paper, including Jes\'us Gallego,
Pablo P\'erez-Gonz\'alez, Joe Helmboldt, and Jarle Brinchmann.  We
would also like to thank the National Optical Astronomy Observatory
(NOAO) for their support, especially the staff of the Cerro Tololo
Inter-American Observatory.

\clearpage

\clearpage

\begin{deluxetable}{l c c c c c c c}
  \tablewidth{0pt}
  \tabletypesize{\small}
  \tablecaption{Local \Halpha\ cosmic star formation rate density measurements\label{t:sfrd}}
  \tablehead{\colhead{Survey} &
             \colhead{Sources} &
             \colhead{$z$} &
             \colhead{log(\SFRDpz)} &
             \colhead{log(\SFRDz)} &
             \colhead{log(\SFRDpo)} &
             \colhead{log(\SFRDo)}}
\startdata
SINGG    & 110 & 0.01 & -2.13 $^{+0.08}_{-0.09}$ & -1.80 $^{+0.13}_{-0.08}$ & -2.14 $^{+0.08}_{-0.09}$ & -1.81 $^{+0.13}_{-0.08}$ \\
UCM (1)  & 264 & 0.02 & -2.45 $\pm$ 0.20 & -1.92 $\pm$ 0.20 & -2.48 $\pm$ 0.20 & -1.95 $\pm$ 0.20 \\
UCM (2)  & 79  & 0.03 & -2.15 $\pm$ 0.11 & -1.60 $\pm$ 0.11 & -2.19 $\pm$ 0.11 & -1.64 $\pm$ 0.11 \\
SDSS (3) & 149660 & 0.10 & \ldots & -1.54 $\pm$ 0.07 & \ldots & -1.66 $\pm$ 0.07 \\
FOCA (4) & 216 & 0.15 & -2.21 $\pm$ 0.15 & -1.82 $\pm$ 0.06 & -2.39 $\pm$ 0.15 & -2.00 $\pm$ 0.06 \\
CFRS (5) & 110 & 0.20 & -1.88 $\pm$ 0.06 & -1.49 $\pm$ 0.06 & -2.12 $\pm$ 0.06 & -1.73 $\pm$ 0.06 \\
\enddata
\tablecomments{Units for columns 4-7 are \SFRDunits; \SFRDpz\ and \SFRDz\ are the SFR density estimates without and with internal dust extinction corrections, respectively, evaluated at the mean redshift $z$ of the survey.  The corresponding $z = 0$ rates are extrapolated assuming \SFRDz\ = \SFRDo\ $(1+z)^{3}$ \\
               1: \cite{b:gallego95} \\
               2: \cite{b:perez03} \\
               3: \cite{b:brinchmann04} \\
               4: \cite{b:sullivan00} \\
               5: \cite{b:tresse98}}
\end{deluxetable}

\begin{deluxetable}{l c c c c c}
  \tablewidth{0pt}
  \tabletypesize{\small}
  \tablecaption{\HI\ mass functions \label{t:himf}}
  \tablehead{\colhead{\HIMF} &
             \colhead{$\alpha$} &
             \colhead{log(\Mstar)} &
             \colhead{$\theta_{\ast}\times10^{3}$} &
             \colhead{log(\SFRD)} &
             \colhead{log(\rhoHI)} \\
             \colhead{(1)} &
             \colhead{(2)} &
             \colhead{(3)} &
             \colhead{(4)} &
             \colhead{(5)} &
             \colhead{(6)}}
\startdata
This paper & -1.41 $\pm$ 0.05 & 9.92 $\pm$ 0.04 & 3.9 $\pm$ 0.7 & -1.80 $^{+0.13}_{-0.08}$ & 7.71 $ \pm 0.03$ \\
\cite{b:zwaan05} & -1.37 $\pm$ 0.03 & 9.86 $\pm$ 0.03 & 4.9 $\pm$ 0.7 & -1.77 $^{+0.15}_{-0.11}$ & 7.73 $^{+0.06}_{-0.08}$ \\
\cite{b:zwaan03} & -1.30 $\pm$ 0.08 & 9.85 $\pm$ 0.06 & 7.5 $\pm$ 1.7 & -1.62 $^{+0.13}_{-0.08}$ & 7.88 $ \pm 0.02$ \\
\cite{b:rosenberg02} & -1.53 & 9.94 & 4.72 & -1.58 $^{+0.14}_{-0.10}$ & 7.91 $ \pm 0.01$ \\
\enddata
\tablecomments{Column descriptions [units]: (1) Source reference. (2) Schechter fit power-law constant. (3) Schechter fit characteristic HI mass [\Msun]. (4) Schechter fit normalization [Mpc$^{-3}$ dex$^{-1}$]. (5) Star formation rate density [\Msun\ yr$^{-1}$ Mpc$^{-3}$]. (6) \HI\ mass density [\Msun\ Mpc$^{-3}$].}
\end{deluxetable}

\begin{deluxetable}{l c c c c}
  \tablewidth{0pt}
  \tabletypesize{\small}
  \tablecaption{Cosmic star formation as a function of mass \label{t:bindata}}
  \tablehead{\colhead{log(\MHI /\Msun )} &
             \colhead{N} &
             \colhead{\tgas} &
             \colhead{\SFRD\ per log(\MHI /\Msun )} &
             \colhead{$l_R$\ per log(\MHI /\Msun )}\\
             \colhead{(1)} &
             \colhead{(2)} &
             \colhead{(3)} &
             \colhead{(4)} &
             \colhead{(5)}}
\startdata
7.0 -- 8.0   & 4  & 5.9 $\pm$ 0.5 & (1.68 $\pm$ 1.31)$\times 10^{-3}$ & (5.53 $\pm$ 3.06)$\times 10^{36}$ \\
8.0 -- 8.5   & 13 & 6.1 $\pm$ 0.4 & (2.88 $\pm$ 3.23)$\times 10^{-3}$ & (5.15 $\pm$ 2.69)$\times 10^{36}$ \\
8.5 -- 9.0   & 20 & 21.4 $\pm$ 3.1 & (1.41 $\pm$ 0.61)$\times 10^{-3}$ & (5.66 $\pm$ 2.52)$\times 10^{36}$ \\
9.0 -- 9.5   & 18 & 9.2 $\pm$ 2.3 & (5.96 $\pm$ 1.38)$\times 10^{-3}$ & (41.9 $\pm$ 33.1)$\times 10^{36}$ \\
9.5 -- 10.0  & 21 & 6.6 $\pm$ 1.1 & (10.2 $\pm$ \ 2.3)$\times 10^{-3}$ & (40.2 $\pm$ \ 8.2)$\times 10^{36}$ \\
10.0 -- 10.5 & 14 & 5.9 $\pm$ 1.0 & (7.28 $\pm$ 1.98)$\times 10^{-3}$ & (32.5 $\pm$ \ 5.2)$\times 10^{36}$ \\
10.5 -- 11.0 & 2  & 6.9 $\pm$ 0.7 & (0.71 $\pm$ 0.17)$\times 10^{-3}$ & (2.48 $\pm$ 0.13)$\times 10^{36}$ \\
\enddata
\tablecomments{Column descriptions [units]: (1) \HI\ mass range. (2) Number of galaxies within \HI\ mass range. (3) Gas cycling timescale [Gyr]. (4) Star formation rate density contribution per decade of \HI\ mass [\Msun\ yr$^{-1}$ Mpc$^{-3}$ dex$^{-1}$]. (5) $R$ band luminosity density contribution per decade of \HI\ mass [erg\ s$^{-1}$ \ang$^{-1}$ Mpc$^{-3}$ dex$^{-1}$].}
\end{deluxetable}

\begin{deluxetable}{l l c c c c}
  \tablewidth{0pt}
  \tabletypesize{\small}
  \tablecaption{Error budget for luminosity densities \label{t:error}}
  \tablehead{\colhead{Uncertainty} &
             \colhead{} &
             \multicolumn{4}{c}{Uncertainty in log(luminosity density)} \\ & &
             \colhead{$l_R'$} &
             \colhead{$l_R$} &
             \colhead{\lhap} &
             \colhead{\lha}}
\startdata
\sl Random Errors \rm    & & & & & \\
\HI\ mass function       & & $\pm$0.029 & $\pm$0.027 & $\pm$0.034 & $\pm$0.029 \\
Sampling                 & & $\pm$0.062 & $\pm$0.073 & $\pm$0.071 & $\pm$0.061 \\
Sky subtraction          & & $\pm$0.001 & $\pm$0.001 & $\pm$0.002 & $\pm$0.002 \\
Continuum subtraction    & & \ldots & \ldots & $\pm$0.010 & $\pm$0.015 \\
Flux calibration         & & $\pm$0.008 & $\pm$0.008 & $\pm$0.013 & $\pm$0.012 \\
\NII\ correction         & & \ldots & \ldots & $^{+0.004}_{-0.006}$ & $^{+0.005}_{-0.008}$ \\
Internal dust extinction & & \ldots & $^{+0.053}_{-0.013}$ & \ldots & $^{+0.107}_{-0.016}$ \\
\bf~~Total Random \rm    & & $\pm$0.069 & $^{+0.094}_{-0.080}$ & $^{+0.078}_{-0.083}$ & $^{+0.127}_{-0.075}$ \\
                         & & & & & \\
\sl Systematic Errors\rm & & & & & \\
\NII\ zeropoint          & & \ldots & \ldots & $\pm$0.002 & $\pm$0.002 \\
Internal dust zeropoint  & & \ldots & $\pm$0.003 & \ldots & $\pm$0.006 \\
Distance model           & & $\pm$0.014 & $\pm$0.017 & $\pm$0.028 & $\pm$0.033 \\
\bf~~Total Systematic\rm & & $\pm$0.014 & $\pm$0.017 & $\pm$0.028 & $\pm$0.033 \\
\enddata
\end{deluxetable}

\begin{deluxetable}{l c c c c c}
  \tablewidth{0pt}
  \tablecaption{Breakdown of \lha\ and $l_R$ by galaxy parameters\label{t:morestat}}
  \tabletypesize{\small}
  \tablehead{\colhead{Percentile} &
             \colhead{10\%} &
             \colhead{25\%} &
             \colhead{50\%} &
             \colhead{75\%} &
             \colhead{90\%}\\
             {} &
             {($R$/\Halpha)} &
             {($R$/\Halpha)} &
             {($R$/\Halpha)} &
             {($R$/\Halpha)} &
             {($R$/\Halpha)}}
\startdata
log(\MHI\ [\Msun]) & 8.19/7.96 & 9.25/9.08 & 9.52/9.58 & 9.97/9.97 & 10.13/10.12 \\
log(\Mdyn\ [\Msun]) & 9.55/9.27 & 10.34/9.87 & 11.07/10.67 & 11.42/11.16 & 11.45/11.42 \\
M$_{R}$ [ABmag] & $-16.86$/$-16.63$ & $-19.64$/$-18.40$ & $-21.68$/$-20.77$ & $-22.49$/$-21.84$ & $-22.69$/$-22.42$ \\
log(SFR [\Msun\ yr$^{-1}$]) & $-1.19$/$-1.05$ & $-0.27$/$-0.33$ & 0.19/0.17 & 0.45/0.74 & 0.89/1.04 \\
A($\Halpha$) [ABmag]& 0.34/0.32 & 0.66/0.51 & 1.11/0.92 & 1.38/1.18 & 1.45/1.36 \\
log($r_{e}(R)$ [pc]) & 3.01/2.88 & 3.47/3.37 & 3.61/3.61 & 3.73/3.73 & 3.89/3.88 \\
log($r_{e}$(\Halpha) [pc]) & 2.90/2.73 & 3.42/3.17 & 3.69/3.54 & 3.84/3.83 & 4.07/3.99 \\
log($r_{e}$(\Halpha)/$r_{e}(R)$) & $-0.19$/$-0.49$ & $-0.06$/$-0.11$ & 0.06/0.02 & 0.14/0.09 & 0.22/0.14 \\
$r_{90}(R)$/$r_{e}(R)$ & 1.96/1.89 & 2.06/1.99 & 2.47/2.23 & 3.26/2.91 & 3.74/3.29 \\
log($EW_{50}$(\Halpha) [\ang]) & 0.53/1.22 & 1.05/1.43 & 1.40/1.64 & 1.64/1.87 & 1.87/1.97 \\
$\mu_{e}(R)$ [$^\dag$] & 22.26/22.28 & 21.42/21.44 & 20.35/20.43 & 19.35/19.97 & 19.04/19.33 \\
log($\Sigma_{e}$(\Halpha) [$^\ddag$]) & $-2.57$/$-2.55$ & $-2.31$/$-2.28$ & $-2.19$/$-1.72$ & $-1.58$/$-1.28$ & $-1.02$/$-0.68$ \\
\enddata
\tablecomments{\\$^\dag$: units are [ABmag arcsec$^{-2}$] \\
               $^\ddag$: units are [\Msun\ yr$^{-1}$ kpc$^{-2}$]}
\end{deluxetable}

\end{document}